# Are conductance plateaus independent events in atomic point contact measurements ? A statistical approach.

**Thomas Leoni, Remi Zoubkoff, Sabrina Homri, Nadine Candoni, Petar Vidakovic, Alain Ranguis, Hubert Klein, Andrés Saúl and Philippe Dumas[1]**
CINaM - CNRS 3118, Aix-Marseille Université, Faculté de Luminy, Case 913, 13288, Marseille Cedex

E-mail: dumas@cinam.univ-mrs.fr

**Abstract.** Conductance-elongation curves of gold atomic wires are measured using a Scanning Tunneling Microscope Break Junction technique at room temperature. Landauer's conductance plateaus are individually identified and statistically analyzed. Both the probabilities to observe, and the lengths of the two last plateaus (at conductance values close to $2e^2/h$ and $4e^2/h$) are studied. All results converge to show that the occurrences of these two conductance plateaus on a conductance-elongation curve are statistically independent events.

## 1. Introduction

Nanowires are now considered to be serious candidates for developing future generations of electronic devices. Decreasing the size of objects has revealed new physical phenomena related to electronic, optical, catalytic, thermal and other properties. The understanding of their structure, stability, and electron transport properties is of paramount importance. At the lower size limit such a nanowire can be made of a single molecular chain [1]. Electrical characterization requires to mastering the electrodes and to understand their properties at the molecular scale.

The experimental setup needed to observe Landauer [2] conductance plateaus while breaking noble metals contacts is quite simple. It is at the level of an undergraduate laboratory experiment: two metallic wires (Pt is fine), a 1.5V battery, a series resistor and an oscilloscope with trace memory are enough to observe the phenomena [3]. The two wires are made to vibrate so they get in and out of contact and therefore, the conductance trace exhibits plateaus. These plateaus correspond to ballistic electron transport through nanometric-sized constrictions. It is nevertheless known that all trials are not successful.

However, to get good statistics, thousands of repeated experiments with data acquisition capabilities are preferred to build a conductance histogram [4,5]. Figure 1 shows such a histogram based on our own measurements displayed together with two examples of typical conductance-elongation curves.

This type of histogram, widely used in the field, shows that some preferred values of conductance are observed. In the case of gold, these preferred values were shown [6] to be approximately integer multiples of the Landauer conductance quantum $G_0 = 2e^2/h$ where $h$ is Planck's constant and $e$ is the electron charge.

Mechanically Controlled Break Junction (MCBJ) or Scanning Tunneling Microscopy operated in the Break Junction regime (STM-BJ) are first choice experimental methods for such conductance studies of metallic nanoconstrictions. Since the early nineties, simultaneously with an extensive theoretical work [7], MCBJ and STM-BJ have been used in air [5], in liquids [8,9] or in vacuum [4,10]. Direct visualization of the nanocontact has been achieved in transmission electron microscopy [11,12]. Conductive-AFM has allowed simultaneous measurement of the stiffness and of the conductance of the wire [13]. Electron-phonon interaction [14] and light emission [15] from such nanocontacts have also been reported. Since the beginning, most of the work has been devoted to metallic junctions [7]. Recently, a new branch has rapidly grown since the publication of the first STM-BJ measurements of single molecule conductance [16].

---

[1] Author to whom any correspondence should be addressed

Are conductance plateaus independent events in atomic point contact measurements ? A statistical approach.

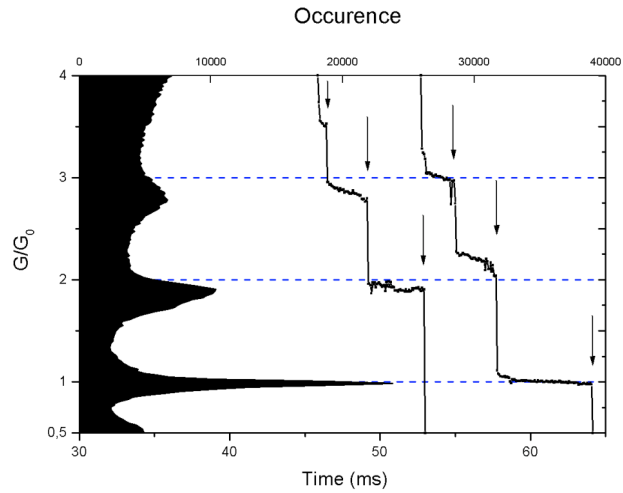

**Figure 1.** Scaled conductance (vertical axis) cumulated histogram (upper horizontal axis) constructed from 9876 gold break junction curves showing the characteristic peaks close to integer multiples of the Landauer conductance quantum $G_0=2e^2/h$. The longer a plateau the higher its contribution to the cumulative histogram. Two conductance elongation (bottom horizontal axis, stretching rate is 100 nm.s$^{-1}$) curves are also shown to illustrate the appearance of the plateaus. Room temperature data acquired with an STM-BJ at 70mV bias.

Conductance histograms exhibiting the above mentioned characteristic preferential peaks have been obtained both at cryogenic temperatures and at room temperature [17]. While being a powerful statistical tool, such histogram analysis (figure 1) masks subtleties of the data, discarding and hiding important information. For example, it is impossible to detect from a histogram analysis, whether the conductance peak at $G_0$ is due to $N$ plateaus of average length $L$ or due to $N/2$ plateaus of average length $2L$.

## 2. Experiments
In this article, we report the development of a new algorithm aimed at statistically analyzing quantities such as the occurrence, the average conductance or the length of individual plateaus. To achieve this, STM-BJ data are first analyzed to identify individual conductance plateaus, separated by abrupt changes of the conductance.

Data were obtained with two home-made and one commercial STM. Despite significant technical differences of the instruments (vibration isolation, drift, stretching speeds) no significant difference was observed in the results. STM were operated at room temperature in ultra-high vacuum ($10^{-10}$ mbar range). The commercial instrument (Omicron VT) was customized to meet the specific requirements of the experiment. The high gain in-vacuum I/V converter was short-circuited and replaced by a current limiting resistor (1kΩ) followed by a low-gain I/V converter ($2.10^5$ V.A$^{-1}$).

The samples are freshly cleaved mica sheets covered by a 100nm layer of gold with a preferential [111] orientation. Gold (99,99% purity) is deposited on heated mica sheets by electron beam evaporation under ultra high vacuum conditions. More details on gold deposition conditions and resulting properties of gold films can be found elsewhere [18]. After rapid transfer in air, samples are introduced in the STM chamber. Au (or Au-covered, W or Pt-Ir) metallic tips are used for the repeated nano-indentation process.

Break junction data are generated and gathered by an independent dedicated software running on a separate computer driving an input/output acquisition card. Briefly, while the STM regulation loop is maintained off, a saw tooth ramp is summed to the (constant) output of the regulation loop to bring the tip into contact prior breaking of the junction by pulling the tip away. Conductance data are simultaneously acquired at a rate of 14kHz. Typically, this sequence lasts few hundreds of milliseconds. Then the regulation loop is active again for ~1s. This cycle is repeated several thousands of times. While data are being acquired, the software displays each conductance-time curve and updates the above-mentioned conductance histogram (see figure1). More time-consuming processing, required for advanced statistics, is performed after acquisition.

Are conductance plateaus independent events in atomic point contact measurements ? A statistical approach.

The first task of the data processing is to identify the conductance plateaus. Conventionally, a plateau is defined as the subset of a conductance elongation curve whose conductance lies between two arbitrary values [19,20]. However this approach has limits and we prefer to define a plateau as a series of consecutive points (at least 3) limited by two *abrupt changes* in the conductance curve.

This criteria is quite natural while it corresponds to what our eyes and brains naturally do to identify a plateau. Indeed when looking at conductance-elongation curves of figure 1 the *abrupt changes* in conductance (marked by downwards arrows) are blindingly obvious. At the contrary, it would be difficult to claim that the values of plateaus' conductance are constant and always close to integer multiple of $2e^2/h$ [19,21].

Our criteria has the second advantage of not introducing the arbitrariness of an "expected" conductance. It is not of prime importance for this study while we do not discuss here the case of molecules of unknown conductance channel transmission [22] but the case of gold atomic junctions, known to have a channel transmission close to one. Our criteria sets two thresholds. Abrupt changes in conductivity are associated with a conductivity change larger than *0.15 $G_0$* within one time step. This corresponds to a decay rate faster than the tunneling decay rate as observed when operating with a STM-BJ or with a MCBJ at high stretching speeds [23]. Plateaus shorter than three consecutive acquisition points (~200 μs) are discarded. We have checked that our results are not sensitive to variations of these thresholds.

As a result of this preliminary analysis, each conductance-elongation curve could be associated with a discrete set of plateaus. Each of these plateaus could then be characterized in terms of average conductance, length, slope, fluctuations, etc. Finally, the data could be used for the statistical analysis described below.

## 3. Results and discussion

We are now armed to analyze the results with respect to occurrence or absence of different plateaus and their eventual correlations.

Figure 2 shows two conductance histograms. In addition to the conductance histogram (in grey) of a set of 9876 conductance-elongation curves, the conductance histogram (in black) of the subset of curves in which a plateau of average conductance between *0.5$G_0$* and *1.5$G_0$* is plotted. As expected, the *1.$G_0$* peaks of both histograms are superimposed. This demonstrates that our analyzing software identifies correctly the events we are interested in. It is however noteworthy to observe that the amplitudes of the second peak (near *2.$G_0$*) of both histograms differ significantly. This means that a significant number of conductance-elongation curves exhibit a plateau around *2.$G_0$* but not around $G_0$. Conductance-elongation curves of figure 1 are typical. They were however selected to show this effect. We will come back to that point later. While our algorithm conveniently sorts the curves, we can now consider more advanced plateaus' statistics.

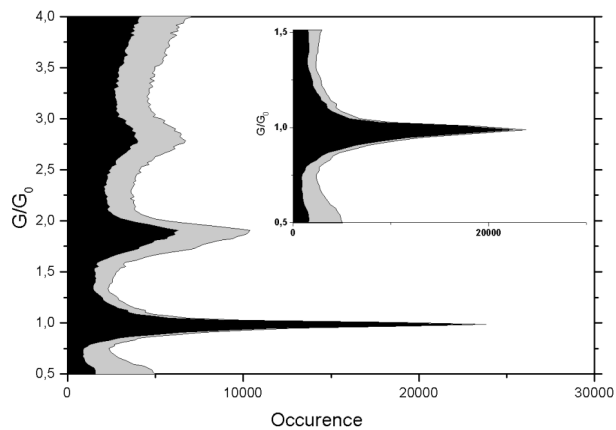

**Figure 2.** Conductance histograms of the entire set (grey) of data and of the subset (black) of the conductance-elongation curves in which a plateau around *1.$G_0$* was identified. The insert is a zoom of the *1.$G_0$* peak for which the two histograms are similar. The small discrepancy clearly visible in the tails of the *1.$G_0$* peak (but also present at the peak maximum) is mainly due to conductance-elongation curves not exhibiting

Are conductance plateaus independent events in atomic point contact measurements ? A statistical approach.

plateaus between $1.5G_0$ and $0.5G_0$. Such curves do not contribute to the conductance histogram of the subset of data. Note that histograms significantly differ around $2.G_0$.

Instead of showing the "standard" conductance histogram, Figure 3 shows the histogram of the *mean conductance* value of the plateaus previously identified. With such a histogram, the length of the plateaus is disregarded. A long or a short plateau with the same average conductance will equally contribute to this histogram. It can be seen that the two types of histogram display similar shapes. This shows that conductance plateaus at $1G_0$ are not longer than conductance plateaus at $1.1G_0$. They are only more frequently observed.

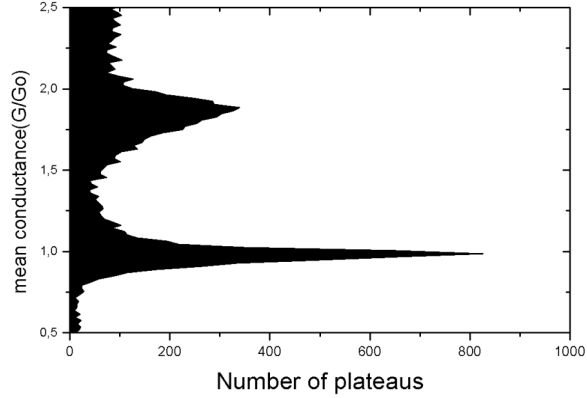

**Figure 3.** Mean conductance histogram of the plateaus identified by the algorithm described in the text.

As it was stressed above, we now focus on the probabilities to obtain a given plateau within a large set of conductance-elongation curves. Coming back to figure 2, we can now stress that the subset of conductance curves containing a plateau around $1G_0$ represents 61% of the complete data set containing nearly $10^4$ curves. Let us note this probability $P_1$. Does this mean that we have 6 chances out of ten to obtain a plateau around $1G_0$ when we perform a conductance curve? Not quite. Indeed, although we found a similar value (63%) for a dataset (line 9 of table 1) of more than $10^5$ curves, we have observed fluctuations of this $P_1$ value computed from different datasets (see table 1). These fluctuations are the result of the correlation between consecutive conductance-elongation curves. If consecutive conductance-elongation curves were uncorrelated or statistically independent, $P_1$ computed from different datasets should be very close to each other provided each dataset is large enough. To probe the correlation between consecutive conductance-elongation curves, the set of 9876 curves (line 1 of table 1) was separated in two halves. From the first 4938 curves (line 3 of table 1) a $P_1$ value of 56% is computed while from the second half (line 4 of table 1) a $P_1$ value of 65% is computed. The net difference between these two values is very large. It corresponds to 8 times the standard deviation. In other words, we cannot assign a $1/\sqrt{N_{curves}}$ error bar to $P_1$. It is why we do not discuss in this article the *absolute* value of $P_1$ as some other groups did [12], but the *conditional* probabilities which we consider more reliable.

We now show that *conditional* probabilities which were neglected in the literature contain valuable statistical information. Table 1 shows, for different sets of data, the probability $P_1$ ($P_2$) to observe a plateau around $1G_0$ ($2.G_0$) together with $P_{1\&2}$, the probability to observe both plateaus on the same conductance-elongation curve.

The last column contains the product $P_1.P_2$. The similarity between $P_{1\&2}$ and $P_1.P_2$ is striking for all the datasets we have investigated. $P_{1\&2} \sim P_1.P_2$ shows that the observation of plateau 1 and the observation of plateau 2 are independent events. When retracting the tip to break the contact, the probability to observe a plateau near $1.G_0$ is the same, whether we have observed a plateau around $2.G_0$ or not. The breaking mechanism is often described as a sequential breaking of conductance channels by pulling away the contacts until the tunnel regime is reached [24]. In such a case, a correlation between the occurrence of consecutive plateaus is expected. However, this correlation is not observed here.

**Table 1.** Comparison of $P_{1\&2}$ (column 4) and $P_1.P_2$ (column 5). Different lines correspond to different experimental conditions (bias, STM, speed, ...) or number of conductance elongation curves ($N_{curves}$, column 1) used for the statistical analysis. Data of line 9 were merged regardless of the experimental conditions during acquisition. All the other figures of this article were plotted with the data corresponding to the first line of this table.

| Line | $N_{curves}$ | $P_1$ | $P_2$ | $P_{1\&2}$ | $P_1.P_2$ |
| --- | --- | --- | --- | --- | --- |
| 1 | 9876 | 61% | 65% | 40% | 39% |
| 2 | 4938 | 61% | 64% | 41% | 39% |
| 3 | 4938 | 56% | 59% | 35% | 33% |
| 4 | 4938 | 65% | 70% | 46% | 46% |
| 5 | 42775 | 54% | 31% | 17% | 17% |
| 6 | 42721 | 63% | 44% | 28% | 28% |
| 7 | 19356 | 86% | 70% | 61% | 61% |
| 8 | 1250 | 62% | 66% | 43% | 41% |
| 9 | 114728 | 63% | 45% | 31% | 29% |

Another result supports this conclusion. Since, in our analysis, the plateaus are characterized by their average conductance and lengths, the algorithm is adapted to sort the lengths of different plateaus separately. The fact that the lengths of the plateaus are related to the stability of atomic wires may serve, in principle, as an additional insight into complicated processes governing the breaking mechanisms. Taking advantage of the superior mechanical stability of a MCBJ design, a recent publication explores these effects for a wide range of junction stretching speeds [25]. Our experiments, like most STM-BJ experiments, correspond to stretching speeds in the high-speed regime. This is also the regime for which the length of the plateaus is speed independent. The histogram built for the occurrence of different lengths of the plateaus around $1G_0$ is shown on Figure 4a. An exponential decay curve fits the data. The same procedure can be applied to the plateaus around $2.G_0$. The length histogram is plotted on figure 4b, together with the length histogram of the $1.G_0$ plateaus.

It is clear that both histograms follow almost indistinguishably the same exponential decay function. Other reported room temperature length histograms exhibit a decay dependence of the probability versus the length [19,20,25]. We propose a simple mathematical model that accounts for the exponential decay reported here. The main aim of this simple model is to provide quantitative parameters to characterize and compare the length histograms. It does not claim to describe the physics of the breaking.

Let $p$ be the probability for an atomic wire to break during a time interval $dt$. The simplest zero-order approximation that can be made is to assume $p$ is proportional only to $dt$. We thus write:

$$p = \frac{dt}{\tau} \quad (1)$$

Let $P(t)$ be the probability to measure a single monoatomic wire during a time t (proportional to its length). If the wire still exists at time $t + dt$, this means that the wire was existing at time $t$ and did not break during $dt$. Thus :

$$P(t+dt) = P(t).(1-p) \quad (2)$$

It is straightforward to show that :

$$P(t) = \exp(-t/\tau)/\tau \quad (3)$$

Are conductance plateaus independent events in atomic point contact measurements ? A statistical approach.

which corresponds to the shape of the histograms shown on figure 4.

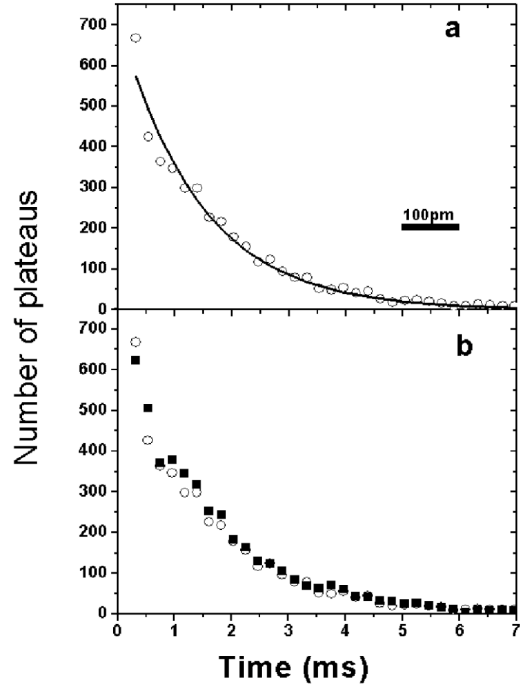

**Figure 4.** (a) Length histogram (circle) of plateau around $1.G_0$. An exponential decay function (solid line) is also plotted. Its time constant is 1.5 ms. (b) Length histogram (squares) of plateau around $2.G_0$ together with length histogram (circles) of plateau 1. The decay constant is the same. The 100pm horizontal bar corresponds to a stretching speed of 100nm/s used. The fact that the two histograms resemble is only due to the coincidence that $P_1 \sim P_2$ on this set of data. In other words the area below the histograms are the same.

Although the main aim of this article is to report quantitative statistical data on the conductance plateaus, a more qualitative discussion of the observed phenomena might be useful as starting point for further research.
We have shown above that the two length histograms exhibit the same decay constant (see figure 4b). A trivial reason why the lifetime of the contacts might be the same for the two plateaus could have been that the lifetime is limited by *external* perturbations, which provide the system enough energy to break the junction. Moreover, random external perturbations would have resulted in a length histogram exhibiting the reported exponential decay form. However, we have checked that the lifetime of the plateaus, is indeed the length of the plateaus. Doubling the elongation rate results in dividing the time constant by a factor of two. The scenario involving an external perturbation can thus be ruled out.
In the range of stretching speeds at which we operated the STM-BJ, it is widely admitted [25] that the $1.G_0$ nanowire breaks when the force to drag one atom from one of the contacts exceeds the maximum tensile strength of the monoatomic wire. This maximum tensile strength is ca. 1.5nN [26]. The monoatomic wire breaks when both contacts have evolved to stiff enough structures.
The length histogram thus reflects the different initial atomic configurations of the contacts rather than the actual length of the wire. The atomic configurations of the $2.G_0$ nanowires are largely unknown. Computer simulations can provide a valuable guide. We have performed tight-binding molecular dynamics simulations to study the stretching and the breaking of the nanojunction [27, 29]. Fig.5 shows two possible atomic configurations of the early stages when the conductance is close to $2.G_0$. Atomic structure of the wires are similar to the ones published by da Silva and coworkers [30]. The upper part of figure 5 exhibits some similarities with two monoatomic wires in parallel. However the detail of the atomic arrangement in the left /right contacts and in the wire differs (see figure 5). This would suggest a different breaking mechanism for $2.G_0$ and $1.G_0$ nanowires. Nevertheless, a ca. 1.5nN force change has been experimentally observed when switching from $2.G_0$ to $1.G_0$ : the same value as for breaking the $1.G_0$ nanowire [26].

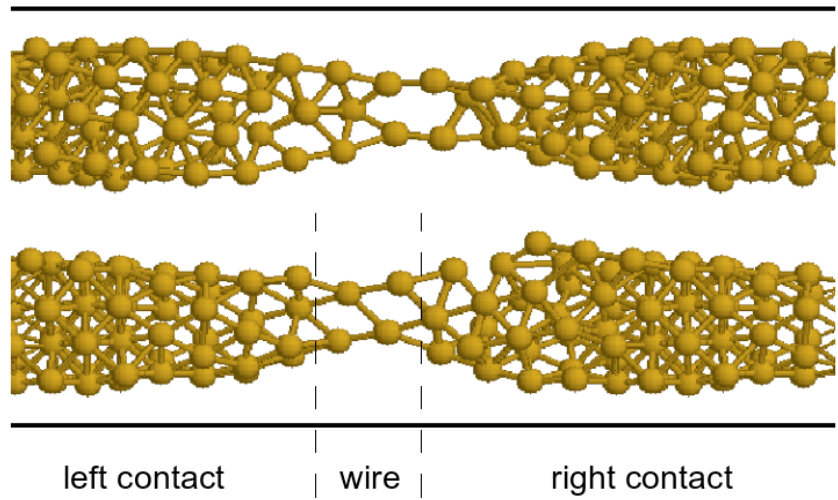

**Figure 5.** Possible atomic configurations for the $2.G_0$ contact calculated by tight-binding molecular dynamics [27].

The breaking of the $2.G_0$ contact thus seems to obey the same physics as the breaking of the monoatomic wire. Indeed, as illustrated in figure 5, the two contacts and the two transmission channels are always asymmetrical. Thus, one of the two conduction channels is expected to break before the other one with the same lifetime as the lifetime of the $1.G_0$ plateau. This could be the explanation for the equality of the two lifetimes.
In such a case, as mentioned above, some correlation between the occurrence of the two plateaus could have been expected. As shown in table 1, this correlation is not observed. It is however important to note that the lifetime of the plateau at $1.G_0$ might be too short-lived to be detected. Following what we explained above for the breaking of the monoatomic nanowire, this is, for instance, likely to occur if the atomic configuration of the two contacts already reaches stiff configurations while stretching the $2.G_0$ nanowire. From an energetic point of view, when the $2.G_0$ nanowire breaks there is a significant elastic energy release. The order of magnitude is 0.15 eV, estimated assuming a tensile force of ca. 1.5nN [26] and a bond elongation of ca. 25pm [26, 29, 30]. This energy release could be enough for a fast, random redistribution of the remaining atoms in the nanojunction. The system could thus evolve either towards the $1.G_0$ contact or directly towards the tunnel regime.

## 4. Conclusion

Conductance-elongation curves of gold nanocontacts were measured with a STM-BJ technique. By detecting the abrupt changes in the conductance, a statistical analysis allows reliable identification of the plateaus. Tens of thousands of curves, acquired at room temperature, with three different STM, were analyzed. Focusing on the two last plateaus, respectively around $2.G_0$ and $1G_0$, we have shown that their occurrences on a conductance curve are *statistically independent events*. Length histograms also support this conclusion.
Although somewhat surprising when imagining the motion of atoms while breaking the contact, this result seems robust and was observed in all our experiments. This effect might be due to the elastic energy release when breaking the $2.G_0$ plateau. It would be interesting to probe this result with independent data acquired with other apparatus especially with MCBJ's in the low speed stretching regime where breaking is dominated by the self-breaking mechanism.

Are conductance plateaus independent events in atomic point contact measurements ? A statistical approach.

**Acknowledgments**
We thank Roger Morin for its interest and stimulating discussions, and Kheya Sengupta for carefully reading the manuscript.